\documentclass[fleqn]{pasj02} 

\usepackage[switch,mathlines]{lineno} 
\usepackage{natbib} 
\usepackage[normalem]{ulem}

\usepackage{booktabs}
\usepackage{bm}
\usepackage{graphicx}
\usepackage{makecell}  
\usepackage{booktabs}
\usepackage[dvipsnames]{xcolor}

\jyear{2026}
\Received{}
\Accepted{}

\begin{document} 
\setlength{\mathindent}{5pt} 

\title{Radiation Pressure Instability-Driven Variability of Line-Driven Disk Winds in AGNs: Connection to Periodic Luminosity Variations and UFO Appearance}

\author{
 Yutaro \textsc{Kuroda},\altaffilmark{1}\email{kuroday@ccs.tsukuba.ac.jp}
 Ken \textsc{Ohsuga},\altaffilmark{1}
 Mariko \textsc{Nomura},\altaffilmark{2}
 and 
 Kenya \textsc{Watarai}\altaffilmark{3}
}

\altaffiltext{1}{Center for Computational Sciences, University of Tsukuba, 1-1-1 Ten-nodai, Tsukuba, Ibaraki 305-8577, Japan}
\altaffiltext{2}{Graduate School of Science and Technology, Hirosaki University, 3 Bunkyo-cho, Hirosaki, Aomori 036-8561, Japan}
\altaffiltext{3}{Kanazawa University Senior High School, 1-1-15 Heiwamachi, Kanazawa, Ishikawa 921-8105, Japan}

\maketitle

\begin{abstract}

We investigated the temporal variability of line-driven disk winds induced by 
radiation pressure instability in accretion disks surrounding supermassive black holes.
This is the first study to self-consistently couple one-dimensional hydrodynamic simulations of accretion disks with two-dimensional radiation hydrodynamic simulations of line-driven winds and to analyze their time-dependent evolution. 
Our results show that periodic luminosity oscillations caused by 
radiation pressure instability lead to corresponding changes in both the mass outflow rate and the covering factor of gas expected to be observed as ultra-fast outflows (UFOs), 
with a delay  
comparable to the viscous timescale around the wind base. 
The viewing angles from which UFOs are expected to be observed also change with time. Our results therefore imply that, depending on the viewing angle, UFOs may not be detected even during high-luminosity states, whereas they may be detected even at lower luminosities. These findings are consistent with observations showing that UFOs are not always detected at high luminosities and can sometimes be detected even at lower luminosities.
For a black hole mass of $10^{7.4}M_\odot$, a mass accretion rate of
$1.3L_{\rm Edd}/c^2$ at a radius of about 100 Schwarzschild radii,
a viscosity parameter of $\alpha=0.1$, and a viscosity prescription parameter
of $\mu=0.45$, the mass outflow rate shows a delay of approximately 2--5 yr
relative to the luminosity variation, and the covering factor consequently
reaches a maximum of $\sim 30\%$ from the high-luminosity phase through
the declining phase.
Our results suggest that the intermittent detection of UFOs in luminosity-variable AGNs can be explained by  
radiation pressure instability driven variability of the line-driven disk winds.

\end{abstract}


\section{Introduction}
The blueshifted iron absorption lines are detected through high-resolution X-ray spectroscopic observations of active galactic nuclei (AGNs). To explain these absorption features, gas with an intermediate ionization state needs to be launched at velocities of several to several tens of percent of the speed of light. Such phenomena are called ultra-fast outflows (UFOs) \citep{Tombesi2010}. The typical column density of UFOs is roughly $N_{\rm H} \sim 10^{23}~\mathrm{cm^{-2}}$, and it has been pointed out that UFOs are launched from regions located at approximately $10^2$--$10^4$ Schwarzschild radii away from the central black hole \citep{Tombesi2012a}.
Furthermore, recent XRISM observations reported that UFOs are composed of numerous gas clumps. However, both the acceleration mechanism of UFOs and the formation mechanism of clumps remain unknown. In addition, the relationship between variations in AGN luminosity and UFOs remains under debate. Some studies reported no clear correlation between variations in AGN luminosity and the presence of UFOs \citep{Tombesi2012b, Pounds2025}. 
On the other hand, \citet{Tombesi2010}, \citet{Kosec2020}, \citet{Reeves2023}, \citet{Reeves2024} and \citet{Baldini2024} pointed out that UFOs tend to appear during phases in which the AGN luminosity decreases. Clarifying the causal relationship between variations in AGN luminosity and the occurrence of UFOs is considered crucial for understanding the physics of accretion and outflows.

One of the most plausible mechanisms for launching UFOs is line driving.
Studies of line-driven disk winds have been actively conducted using two-dimensional radiation hydrodynamic simulations. \citet{Proga1998}, \citet{Proga2000} and \citet{Proga2004} revealed that line-driven disk winds with funnel-shaped structures can be launched from regions located at several hundred Schwarzschild radii. In addition, \citet{Nomura2016} and \citet{Nomura2017} demonstrated that line-driven disk winds can roughly reproduce the velocity, column density, and ionization parameter of UFOs estimated from the observations by \citet{Tombesi2011}, as well as the mass outflow rate, momentum, and kinetic energy reported by \citet{Gofford2015}.
\citet{Mizumoto2021} also calculated radiation spectra based on their simulation results and successfully reproduced the X-ray absorption line profiles of UFOs. 
These 
simulations assume a steady accretion disk, which results in quasi-steady disk winds. As a result, 
the relationship between variations in AGN luminosity and the occurrence of UFOs remains unresolved. In addition, 
\citet{Nomura2020} adopted a model in which the mass accretion rate in the disk decreases due to the presence of disk winds. However, their model still focuses on a quasi-steady structure. Magnetically driven disk winds are also considered a plausible origin of UFOs
\citep{Blandford1982,Proga2003,Fukumura2015,Yang2021,Wang2022}. Nevertheless, no studies have focused on the relationship between variations in AGN luminosity and the occurrence of UFOs.

A possible mechanism of variability in luminous accretion disks where line-driven winds can be launched
is considered to be radiation pressure instability \citep{Honma1991,Szuszkiewicz1998,Kato2008,Grzedzielski2017}.
Disk variability caused by radiation pressure instability has successfully explained luminosity variations in several black hole systems, including the microquasar GRS 1915+105 \citep{Watarai2003,Janiuk2005,Wu2016}, intermittent radio activity in AGNs \citep{Czerny2009}, changing-look behavior in NGC~1566, NGC~4151, and NGC~5548 \citep{Sniegowska2020}, and quasi-periodic eruptions in GSN~069 \citep{Pan2022}.
For supermassive black holes,
the timescale of such variability is expected to be on the order of several to several tens of years.
Therefore, it may be related to long-term luminosity variations associated with UFOs.

In this study, we investigate the structure and 
evolution of line-driven disk winds launched from 
accretion disks undergoing variability due to   
radiation pressure instability. 
In this study, we, for the first time, self-consistently couple one-dimensional hydrodynamic simulations of accretion disks with two-dimensional radiation hydrodynamic simulations of line-driven disk winds and investigate their time-dependent evolution.
The disk simulation accounts for mass, momentum, and energy losses caused by the wind, and the wind model incorporates changes in the disk's density distribution and radiation field. By iteratively referencing the results of the disk and wind simulations, we self-consistently examine the coupled structure and evolution of the disk and disk wind. This paper is organized as follows. In Section 2, we describe the computational methods. In Section 3, we present the simulation results. Section 4 is devoted to conclusions and discussions.

\section{Method}
In this study, we simultaneously perform one-dimensional numerical hydrodynamic simulations for the accretion disk and two-dimensional axisymmetric radiation hydrodynamic simulations for the line-driven disk wind. The time evolution of both components is treated in a self-consistent manner. The outline of the simulations is as follows.

In the time-dependent simulation of the accretion disk, we perform one-dimensional hydrodynamic simulations that incorporate the mass, momentum, and energy losses due to disk wind. These loss rates are evaluated from two-dimensional radiation hydrodynamic simulations of line-driven disk winds and are updated every year.
In turn, 
in the two-dimensional radiation hydrodynamic simulations of the disk wind, variations in the surface density and effective temperature of the accretion disk are taken into account. These quantities are obtained from the results of the one-dimensional hydrodynamic simulations of the disk and are also updated every 1 year.
In this way, the surface density and effective temperature of the disk, along with the mass outflow from the disk wind, allow mutual interaction between the two simulations. 
It should be noted that changing the update interval from 1 year to 0.5 or 2 years does not significantly affect the results of this study. This is because 
the timescale of variation in both the disk and the wind is sufficiently longer than 1 year.

\subsection{Method of disk simulation}\label{sec:disk}

We assume that the accretion disk is axisymmetric and in vertical hydrostatic equilibrium and perform one-dimensional hydrodynamic simulations in the radial direction. 
{The basic} equations in cylindrical coordinates ($R, \psi, z$) are given as follows: the equation of continuity,
\begin{gather}
\label{sec:5}  \frac{\partial}{\partial t}(R\Sigma)+\frac{\partial}
{\partial R}(R\Sigma v_{R})=-\frac{s}{2\pi R},
\end{gather}
the radial component of the momentum conservation,
\begin{gather}
\notag    \frac{\partial}{\partial t}(R\Sigma v_{
R})+\frac{\partial}{\partial R}(R\Sigma v_{R}^2+R\Pi)
\\ \label{sec:66}  = \left(1-\frac{d\ln \Omega_{\mathrm{k}}}{d\mathrm{ln}R}\right)\Pi+\Sigma(v_{\psi}^2-(R\Omega_{\mathrm{K}})^2)-\frac{s}{2\pi R}v_{R},
\end{gather}
the angular momentum conservation,
\begin{gather}
\label{sec:7}  \frac{\partial}{\partial t}(R^2\Sigma v_{\psi})+\frac{\partial}{\partial R}(R\Sigma v_{R}v_{\psi}+R^2T_{R\psi})=-\frac{s}{2\pi R}v_{\psi},
\end{gather}
 and the energy equation,
\begin{gather}
\notag         \frac{\partial}{\partial t}(R\Sigma \epsilon_{\mathrm{tot}})+ \frac{\partial}{\partial R}
[R(\Sigma\epsilon_{\mathrm{tot}}+\Pi)v_{R}+RT_{R\psi}v_{\psi}]\\ \label{sec:99} =-2RF^{-}-\frac{s}{2\pi R}\left(\epsilon_{\mathrm{tot}}+\frac{\Pi}{\Sigma}\right).
\end{gather}

Here, $\Sigma$ is the surface density, and $\Pi$ is the vertically integrated total pressure, defined as the sum of the gas pressure and radiation pressure, $p_{\rm gas} + p_{\rm rad}$. $v_{R}$ and $v_{\psi}$ are the radial and azimuthal velocities, respectively. $T_{R\psi}$ denotes the vertically integrated $R$–$\psi$ component of the viscous stress tensor, and $\Omega_{\rm K}$ is the Keplerian angular velocity. $\epsilon_{\mathrm{tot}}$ represents the total specific energy, and $s$ is the mass loss rate per unit radial length due to disk winds (the calculation of $s$ is described in subsection \ref{sec:cal_s}). $F^{-}$ is the radiation cooling flux per unit surface area, for which blackbody radiation is assumed \citep{Ichikawa1992, Kimura2023}.
The viscous stress tensor is given by $t_{r\varphi} = \alpha p_{\mathrm{gas}}^{\mu}(p_{\rm gas}+p_{\rm rad})^{1 - \mu}$, where $\alpha$ is the viscosity parameter and $\mu$ is a free parameter. The vertical structure of the disk is calculated by assuming a polytropic relation, following the method of \citet{Watarai2003}. In this study, we employ the numerical code developed by \citet{Watarai2003}, which we have modified to include mass, momentum, and energy losses due to the disk wind.

The computational domain, $R = [3r_{\mathrm{s}}, 2000r_{\mathrm{s}}]$ (where $r_{\mathrm{s}}$ is the Schwarzschild radius),
is divided into 200 grid points. Radial grids are distributed
uniformly on the logarithmic scale. As the initial condition, we adopt the transonic solution presented in \citet{Matsumoto1984}.
At the inner boundary, the density and pressure in the ghost cells are fixed to very small values. The velocity is set so that its radial gradient near the inner boundary is preserved, which suppresses reflection waves and mimics free outflow. At the outer boundary, all physical quantities in the ghost cells are fixed to their initial values.

At the beginning of the simulation, a strong burst occurs, and its influence persists for a long time in the region at 
$r > 100r_{\rm s}$. The region around $r \sim 100r_{\rm s}$ becomes quasi-steady, 
while the inner region ($r < 100r_{\rm s}$) exhibits periodic variations due to the disk instability. Both disk radiation and disk winds 
primarily originate from the region within $100r_{\rm s}$. 
Therefore, we define the quasi-steady mass accretion rate at $r = 100r_{\rm s}$ as the mass supply rate to the vicinity of the black hole, denoted as $\dot{M}_{\mathrm{sup}}$.
In this study, we use $\dot{M}_{\mathrm{sup}} = 1.3L_{\mathrm{Edd}}/c^2$ as the fiducial value, and also explore cases with $0.32L_{\mathrm{Edd}}/c^2$, $0.56L_{\mathrm{Edd}}/c^2$, 
and $3.8L_{\mathrm{Edd}}/c^2$. Here, $L_{\mathrm{Edd}}$ is the Eddington luminosity and $c$ is the speed of light. The viscosity parameter $\alpha$ is set to $0.1$ as the fiducial value, with an additional case of $\alpha = 0.06$ considered. The parameter $\mu$ is set to 0.45 as the fiducial value. 
The adopted values of $\mu$ are consistent with the range ($0.3$–$0.5$) suggested by studies on luminosity variations in X-ray binaries and active galactic nuclei \citep{Czerny2009, Wu2016, Grzedzielski2017}.
The mass of the central black hole ($M_{\mathrm{BH}}$) is set to $10^{7.4} M_\odot$. 

\subsection{Method of wind simulation}
We perform two-dimensional radiation hydrodynamic simulations using 
spherical coordinates $(r, \theta, \varphi)$, where $\theta = 0^{\circ}$ corresponds to the rotation axis of the disk. We assume axisymmetry about 
the rotation axis. The basic hydrodynamic equations consist 
of the mass conservation, 
\begin{gather}
\label{radhy1}\frac{\partial \rho}{\partial t} + \nabla \cdot (\rho \bm{v}) = 0,
\end{gather}
the momentum conservation,
\begin{gather}
\label{radhy2}\frac{\partial (\rho v_r)}{\partial t} + \nabla \cdot (\rho v_r \bm{v}) = - \frac{\partial p}{\partial r} + \rho \left[ \frac{v_\theta^2}{r} + \frac{v_\varphi^2}{r} + g_r + f_{\text{rad}, r} \right],\\
\label{radhy3}\frac{\partial (\rho v_\theta)}{\partial t} + \nabla \cdot (\rho v_\theta \bm{v}) = - \frac{1}{r}\frac{\partial p}{\partial \theta} + \rho \left[ \frac{v_\theta v_{r}}{r} + \frac{v_\varphi^2}{r}\cot \theta + g_\theta + f_{\text{rad}, \theta} \right],\\
\label{radhy4}\frac{\partial (\rho v_\varphi)}{\partial t} + \nabla \cdot (\rho v_\varphi \bm{v}) = - \rho \left[ \frac{v_\varphi v_r}{r} + \frac{v_\varphi v_\theta}{r} \cot \theta \right],
\end{gather}
and the energy equation,
\begin{gather}
\label{radhy5}\notag \frac{\partial}{\partial t} \left[ \rho \left( \frac{1}{2} v^2 + e \right) \right] + \nabla \cdot \left[ \rho \bm{v} \left( \frac{1}{2} v^2 + e + \frac{p}{\rho} \right) \right] \\ = \rho \bm{v} \cdot \bm{g} +\rho \bm{v}\cdot \bm{f}_{\mathrm{rad}} +\rho \mathcal{L} .
\end{gather}

Here, $\rho$ is the mass density, 
and $\bm{v} = (v_{r}, v_{\theta}, v_{\varphi})$ is the velocity vector. The gas pressure is denoted by $p$, and the specific internal energy $e$ is given by $e = p/\rho(\gamma - 1)$, 
with the adiabatic index set to $\gamma = 5/3$. The gravitational acceleration is represented by $\bm{g} = (g_{r}, g_{\theta})$. Note that the $\theta$-component of gravity is nonzero because the origin of the polar coordinate system is offset from the black hole 
(see below).
The radiation force is given by $\bm{f}_{\mathrm{rad}} = (f_{\mathrm{rad}, r}, f_{\mathrm{rad}, \theta})$, and $\mathcal{L}$ represents the net cooling rate. The cooling/heating processes included in $\mathcal{L}$ are Compton heating/cooling, X-ray photoionization heating, recombination cooling, bremsstrahlung cooling, and line cooling \citep{Proga2000}.

We set up 100 radial and 160 polar grid points, with the computational domain defined as $r = [30r_{\mathrm{s}}, 1500r_{\mathrm{s}}]$ and $\theta = [0^\circ, 90^\circ]$.
It should be noted that the origin of the polar coordinate system is offset by $z_{\mathrm{off}}$ along the rotation axis from the center of the black hole. As a result, the $\theta = 90^\circ$ surface lies above the equatorial plane by a height of $z_{\mathrm{off}}$. In this study, we adopt $z_{\mathrm{off}} = 0.66\left(\dot{M}_{\mathrm{sup}}/{(L_{\mathrm{Edd}}/c^2)}\right)r_{\mathrm{s}}$, 
which corresponds to the scale height at $30r_{\mathrm{s}}$, calculated by applying a polytropic relation $\Sigma \propto \Pi^{1 + 1/N}$ with $N = 3$ to a standard accretion disk \citep{Shakura1973}. The numerical method for computing 
the disk wind in this study is almost the same as that employed in \citet{Nomura2017}. The only differences are in the adopted value of $z_{\mathrm{off}}$, which was $z_{\mathrm{off}} = 0.19\left(\dot{M}_{\mathrm{sup}}/{(L_{\mathrm{Edd}}/c^2)}\right)r_{\mathrm{s}}$ in their study, and in the treatment of the density distribution and effective temperature at the $\theta = 90^\circ$ surface. These modifications are described in detail in the next subsection.

\subsection{Mutual Interaction between the Accretion Disk and the Disk Wind}\label{sec:cal_s}
In this study, the effective temperature distribution of the disk and the gas density at the $\theta = 90^\circ$ surface used in the disk wind simulation 
are derived based on the results of the accretion disk simulation. On the other hand,  
the mass loss rate, which is used
in the disk simulation, is estimated from the disk wind simulation. This approach allows us to self-consistently account for 
the mutual interaction between the disk and the disk wind.

The radial distributions of the gas density and effective temperature at the $\theta = 90^\circ$ surface, which are required in the wind simulation, are calculated for times $t_n < t \leq t_{n+1}$ as follows:
\begin{gather}
  \rho
  =\frac{\Sigma}{H} \mathrm{exp}\left(-\frac{z_{\mathrm{off}}}{H}\right),
\end{gather}
and
\begin{gather}
T_{\mathrm{eff}}
=   \left(\frac{F^{-}}{\sigma}\right)^{\frac{1}{4}}.
\end{gather}
where $\Sigma$ and $F^{-}$ are the surface density and radiation flux of the disk at $t = t_n$ and $R = r$, obtained from the disk simulation described in subsection \ref{sec:disk}. $\sigma$ is the Stefan–Boltzmann constant. 
The scale height $H$ of the disk at $t = t_n$ and $R = r$ is given by
\begin{gather}
    H=\sqrt{(2N+3)\frac{\Pi}{\Sigma}}\frac{1}{\Omega_{\mathrm{K}}},
\end{gather}
with $N = 3$. Since the radial grids of the disk and wind simulations are not aligned, linear interpolation is applied to obtain the corresponding values.

On the other hand, the mass loss rate per unit radius ($s(R_i)$), which is  required at radius $R_i$ and time $t_n < t \leq t_{n+1}$ in the disk simulation, is given by 
\begin{gather}
s(R_i) =  \frac{\overline{\dot{M}}_{\mathrm{out}}}{\sqrt{2\pi\tau^2}} \exp\left[ -\frac{(R_i - R_{\mathrm{launch}})^2}{2\tau^2} \right], \label{eq:13}
\end{gather}
where $\overline{\dot{M}}_{\mathrm{out}}$ is the time-averaged mass outflow rate measured at the outer boundary $r = r_{\mathrm{out}} \equiv 1500\,r_{\mathrm{s}}$, calculated as
\begin{gather}
\overline{\dot{M}}_{\mathrm{out}}=\frac{4\pi r^2}{t_n - t_{n-1}} \int_{t_{n-1}}^{t_n} \int_{0^\circ}^{89^\circ} \rho v_{r} \sin\theta \, d\theta \, dt.
\end{gather}
The launching radius ($R_{\mathrm{launch}}$) is calculated 
using the time-averaged angular momentum outflow rate at $r = r_{\mathrm{out}}$, 
\begin{gather}
\overline{\dot{l}}_{\mathrm{out}}=\frac{4\pi r^3}{t_n-t_{n-1}}\int_{t_{n-1}}^{t_n}\int_{0^\circ}^{89^\circ}\rho  v_{\varphi} v_{r}  \mathrm{sin}^2\theta d\theta dt,
\end{gather}
as
\begin{gather}
R_{\mathrm{launch}} = \frac{1}{GM_{\mathrm{BH}}} \left( \frac{\overline{\dot{l}}_{\mathrm{out}}}{\overline{\dot{M}}_{\mathrm{out}}} \right)^2,  \label{eq:16}
\end{gather}
where $G$ is the gravitational constant.
The width of the mass loss distribution, $\tau$, is determined by minimizing the quantity $\sum_i \left( S(R_i) - s(R_i) \right)^2$, where 
\begin{gather}
S(R_i) = \frac{ \sum_{k} \Delta \overline{\dot{M}}_{\mathrm{out},k}}{R_{i+\frac{1}{2}} - R_{i-\frac{1}{2}}}, \\
\text{for } k \text{ such that } R_{\mathrm{launch},k} \in [R_{i-\frac{1}{2}}, R_{i+\frac{1}{2}}].
\end{gather}
Here, 
$R_{i+\frac{1}{2}}$ and $R_{i-\frac{1}{2}}$ are the radial positions of the cell surfaces surrounding the $i$-th cell.
Each $\Delta \overline{\dot{M}}_{\mathrm{out},k}$ denotes the time-averaged mass outflow rate at $r = r_{\mathrm{out}}$, and is given by
\begin{gather}
\Delta\overline{\dot{M}}_{\mathrm{out},k} = \frac{4\pi r^2}{t_n - t_{n-1}} \int_{t_{n-1}}^{t_n} \int_{\theta_{k-1/2}}^{\theta_{k+1/2}} \rho v_{r} \sin\theta \, d\theta \, dt.
\end{gather}
The launching radius associated with each angular zone ($R_{\mathrm{launch},k} $) is estimated from 
\begin{gather}
R_{\mathrm{launch},k} = \frac{1}{GM_{\mathrm{BH}}} \left( \frac{\Delta \overline{\dot{l}}_{\mathrm{out},k}}{\Delta \overline{\dot{M}}_{\mathrm{out},k}} \right)^2,
\end{gather}
where $\Delta \overline{\dot{l}}_{\mathrm{out},k}$ is the time-averaged angular momentum outflow rate in the same zone, calculated as
\begin{gather}
\Delta \overline{\dot{l}}_{\mathrm{out},k} = \frac{4\pi r^3}{t_n - t_{n-1}} \int_{t_{n-1}}^{t_n} \int_{\theta_{k-1/2}}^{\theta_{k+1/2}} \rho v_{\varphi} v_{r} \sin^2\theta \, d\theta \, dt. \label{eq:21}
\end{gather}
Throughout this paper, we adopt a fixed time interval of $t_n - t_{n-1} = 1~\mathrm{yr}$, which is sufficiently shorter than the characteristic timescale of the disk variability. We confirmed that varying this interval does not significantly affect the simulation results.

\section{Results}
\subsection{Fiducial model}
\begin{figure}[htbp]
  \begin{center}
    \includegraphics[width=0.5\textwidth]{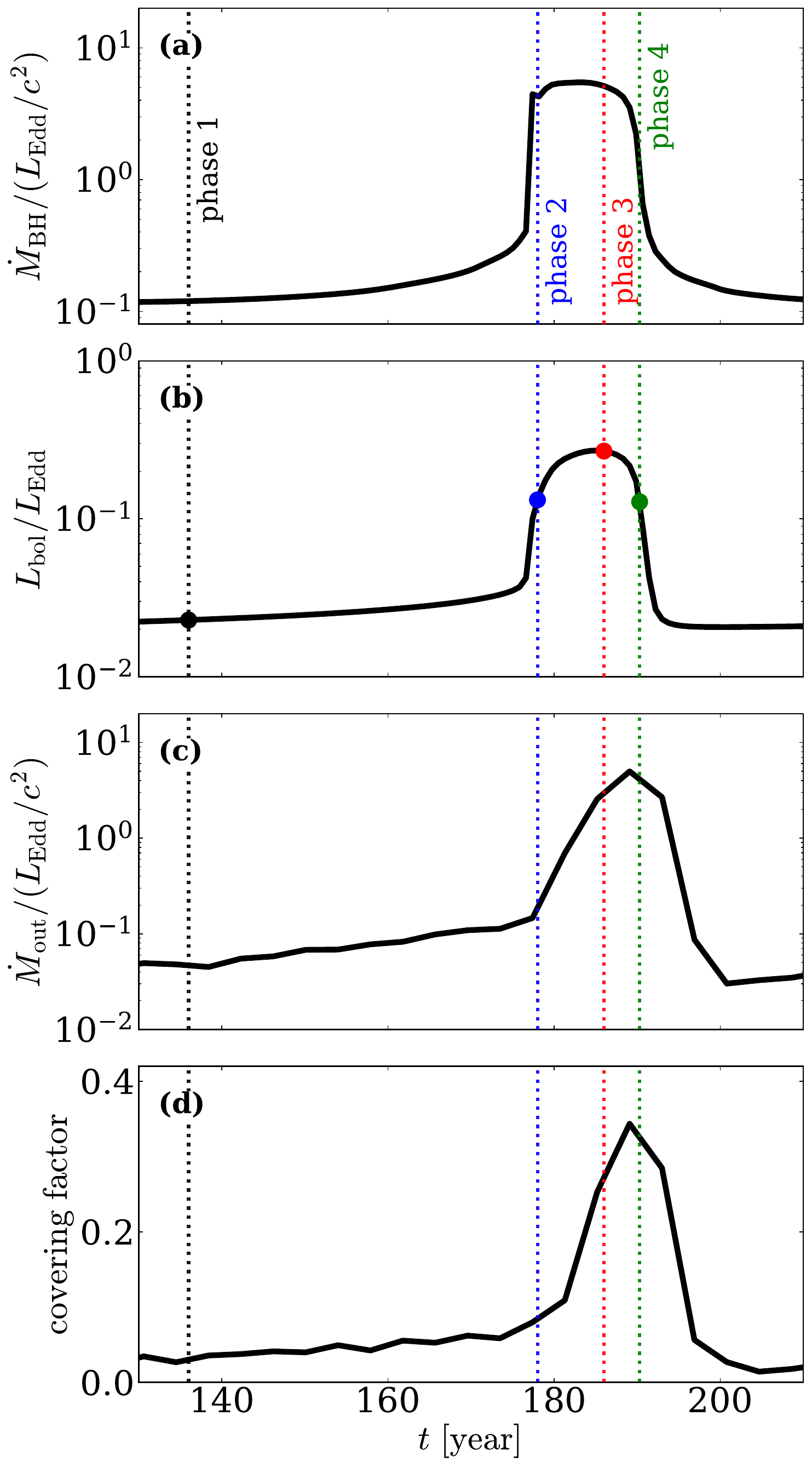}

  \end{center}
    \caption{Time evolution of the mass accretion rate onto the black hole (panel a), the disk luminosity (panel b), the mass outflow rate of the disk wind at $1500r_{\mathrm{s}}$ (panel c), and the covering factor of gas satisfying the UFO criterion (panel d) for the fiducial model.
    The black, blue, red, and green dotted lines mark the selected snapshot epochs in phases 1, 2, 3, and 4, respectively.     
    The mass outflow rate and covering factor
    are temporally averaged over \(\Delta t=5\) yr, whereas the values quoted in the text for individual snapshots are instantaneous and may therefore differ from those shown in the figure.  
    Alt text:
    Multi-panel figure with four line graphs showing, from panels a to d, the time evolution of the accretion rate, disk luminosity, mass outflow rate, and covering factor. 
    The black, blue, red, and green dotted lines indicate the snapshot epochs corresponding to phases 1, 2, 3, and 4, respectively.
    }
    \label{fig:time_fidu}
\end{figure}

The black solid lines in 
Fig.~\ref{fig:time_fidu}
represent the time evolution of the quantities during one representative cycle of the limit-cycle oscillations driven by the disk's radiation pressure instability of the fiducial model (see Table \ref{tablefull}).
The mass accretion rate onto the black hole is expressed as
\begin{gather}
\dot{M}_{\mathrm{BH}} = -2\pi R_{\mathrm{in}} v_{R} \Sigma,
\end{gather}
with $R_{\mathrm{in}} = 3r_{\rm s}$ (panel a). 
The bolometric luminosity of the accretion disk is calculated as
\begin{gather}
L_{\mathrm{bol}} = \int_{R_{\mathrm{in}}}^{R_{\mathrm{out}}} 4\pi R \sigma T_{\mathrm{eff}}^4(R) dR,
\end{gather}
where $R_{\rm out} = 2000r_{\rm s}$ (panel b). 
The mass outflow rate due to the disk wind is obtained from
\begin{gather}
\dot{M}_{\mathrm{out}}=4\pi r^2\int_{0^\circ}^{89^\circ}\rho v_{r} \sin
\theta d\theta, 
\end{gather}
evaluated at $r = r_{\mathrm{out}}$ (panel c). 

Panels~a and b of Fig.~\ref{fig:time_fidu} show that when
$\dot{M}_{\mathrm{BH}}/(L_{\mathrm{Edd}}/c^2)$ increases from
$\sim0.1$ to $\sim5.2$, the luminosity rises from
$\sim0.02L_{\mathrm{Edd}}$ to $\sim0.27L_{\mathrm{Edd}}$.
Corresponding to luminosity variations,
the mass outflow rate also oscillates periodically between $\sim 0.03L_{\mathrm{Edd}}/c^2$ and $\sim 4.7L_{\mathrm{Edd}}/c^2$ (panel c). Although the luminosity varies by about one order of magnitude, the variation in mass outflow rate spans roughly two orders of magnitude, indicating that the disk wind's mass loss rate is highly sensitive to luminosity changes.
The mass outflow rate also lags slightly behind the luminosity
variation, and the physical origin of this time delay is discussed
in Section~3.2.
Time variability in the mass outflow rate causes temporal fluctuations in the 
covering factor of gas satisfying the UFO criterion (panel d). 
We define the UFO criterion as $N_{\rm H} = 10^{22}$–$10^{24}\,{\rm cm^{-2}}$ for gas satisfying $\log\xi = 2.5$–$5.5$ and $v_r > 10^4\,{\rm km\,s^{-1}.}$
Here, the ionization parameter is given by
\begin{gather}
\xi = \frac{f_{\mathrm{X}}L_{\mathrm{bol}}}{n r^2} e^{-\tau_{\mathrm{X}}}\ \mathrm{erg}\ \mathrm{s^{-1}}\ \mathrm{cm}, 
\end{gather}
where $f_{\mathrm{X}} (=0.1)$ is the ratio of the X-ray luminosity to the disk luminosity, $n$ is the number density, $\tau_{\mathrm{X}}$ is the optical depth for the X-ray \citep{Nomura2016}. 
The column density is calculated as
\begin{gather}
N_{\mathrm{H}}= \int_{30r_{\mathrm{s}}}^{r} n(r',\theta)dr'.
\end{gather}
Although our definition of the UFO criterion is slightly different from the one adopted in \cite{Nomura2016}, our conclusions are unaffected.
The covering factor reaches about 0.35 during the high-luminosity phase, but falls to as low as 0.02 in the low-
luminosity phase. It should be noted, however, that the mass outflow rate and covering factor lag behind the luminosity variations. We discuss the origin of this lag below. Time-averaged values of mass outflow rate, disk luminosity, mass accretion rate, and covering factor are summarized in Table \ref{tablefull}. Even when the disk evolution is calculated without including the mass, momentum, and energy losses due to disk winds, 
the disk still oscillates with nearly the same period, indicating that the presence of winds does not significantly affect the period. However, due to the wind-driven mass loss, the time-averaged mass accretion rate (and corresponding luminosity) is reduced by about 66 \% (68 \%)

\begin{figure}[htbp]
  \begin{center}
    \includegraphics[width=0.5\textwidth]{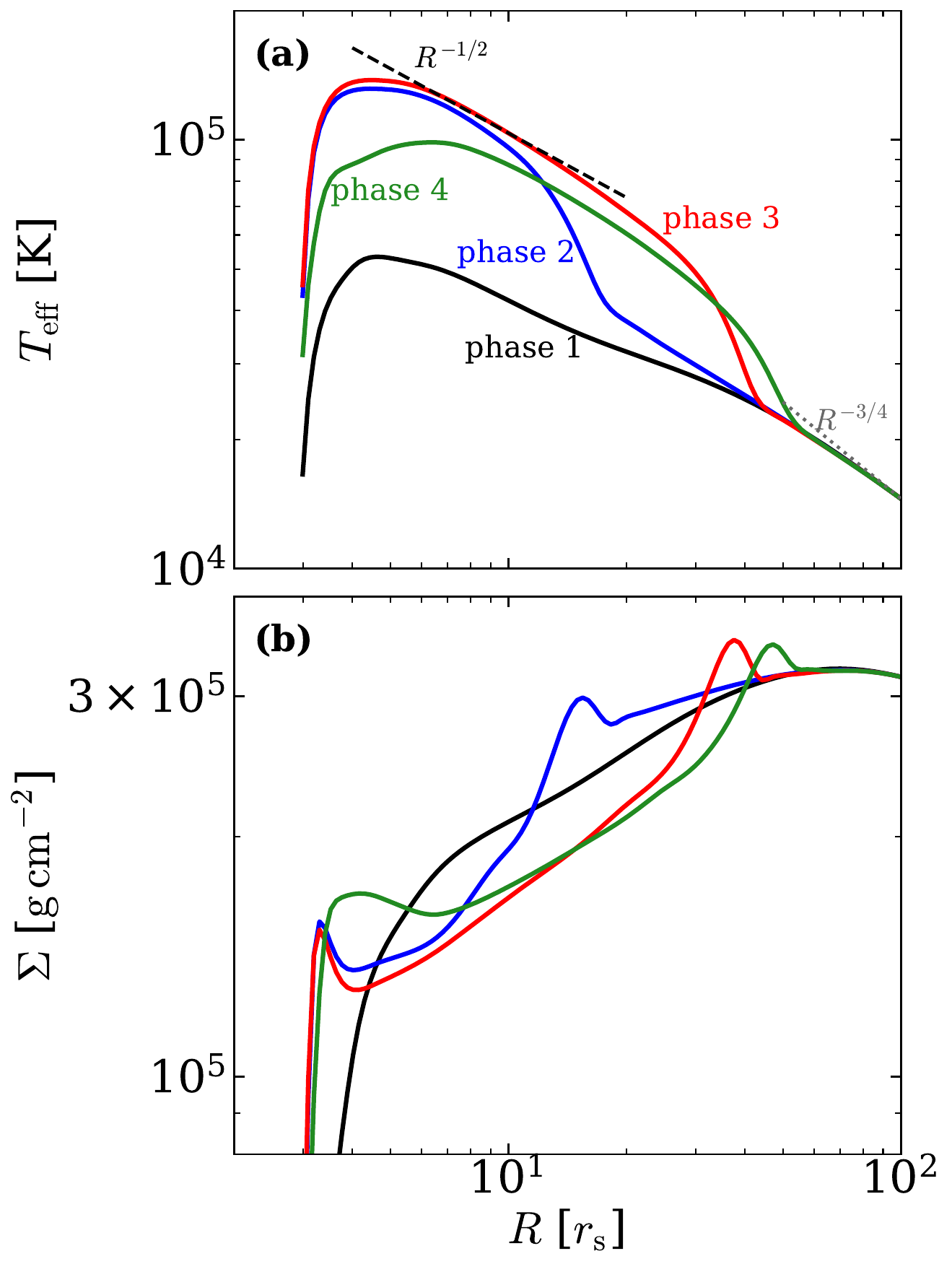}

  \end{center}
    \caption{
    Radial profiles 
    of the effective temperature (panel a) and surface density (panel b) of the accretion disk for the fiducial model. The black, blue, red, and green lines correspond to
phases 1, 2, 3, and 4, respectively. Alt text: A line graph showing radial
profiles of the effective temperature and surface density of the disk for four phases of the cycle,
plotted as functions of radius with four curves representing the different
phases.
    }
    \label{fig:teff_sigma} 
\end{figure}

To investigate the evolution of the disk structure over one cycle of the periodic variability, Figure \ref{fig:teff_sigma} displays the radial profiles of the effective temperature and surface density.
We here define four representative phases in one cycle of the periodic variability: phase 1, a low-luminosity state ($L \lesssim 0.1L_{\mathrm{Edd}}$, $t = 136$ yr); phase 2, a rising stage ($L\sim 0.1L_{\mathrm{Edd}}$, $t = 178$ yr); phase 3, a high-luminosity state ($L \approx 0.27L_{\mathrm{Edd}}$, $t = 185$ yr); and phase 4, a declining stage ($L \sim 0.1L_{\mathrm{Edd}}$, $t = 190$ yr).
These phases are marked by black, blue, red, and green filled circles in panel b, with the corresponding vertical dashed lines shown in all panels.
In all phases, the mass supply rate at the outer boundary remains
constant, and thus the structure of the outer disk changes little.
Around $R\sim100r_{\rm s}$, the effective-temperature profile is
slightly flatter than the $T_{\rm eff}\propto R^{-3/4}$ relation
expected for a standard disk because of mass loss through the disk
wind. 
A high-temperature region develops in the inner
disk owing to the radiation pressure instability. It first appears
near the black hole in phase~2 and expands outward in phase~3. As a
result, the effective-temperature profile in the inner disk approaches
$T_{\rm eff}\propto R^{-1/2}$, which is characteristic of a slim disk.
The surface density decreases during this evolution. In phase~4, the
high-temperature region begins to cool, and the disk structure
gradually returns to the low-luminosity state of phase~1. This cycle
repeats periodically. 
The behaviors of the effective temperature and
surface density are characteristic of limit-cycle oscillations \citep{Czerny2009, Wu2016, Grzedzielski2017}.
These periodic changes in the disk structure give rise to the time variability of the line-driven disk wind (see Section~3.2).

\begin{table*}[htbp]
\centering

\setlength{\tabcolsep}{10pt}
\renewcommand{\arraystretch}{1.45}

\resizebox{\textwidth}{!}{%
\begin{tabular}{lccccccccc}
\toprule

Model
&
$\alpha$
&
$\dfrac{\dot{M}_{\mathrm{sup}}}{L_{\mathrm{Edd}}/c^2}$
&
$\dfrac{\dot{M}_{\mathrm{BH}}}{L_{\mathrm{Edd}}/c^2}$
&
$\dfrac{L_{\mathrm{bol}}}{L_{\mathrm{Edd}}}$
&
$\dfrac{\dot{M}_{\mathrm{out}}}{L_{\mathrm{Edd}}/c^2}$
&
\makecell{covering\\factor}
&
\multicolumn{3}{c}{Variability timescales} \\

\cmidrule(lr){8-10}

&
&
&
&
&
&
&
\makecell{$t_{\mathrm{high}}$\\{[yr]}}
&
\makecell{$P_{\mathrm{cyc}}$\\{[yr]}}
&
\makecell{$t_{\mathrm{delay}}$\\{[yr]}} \\

\midrule
\addlinespace[3pt]

Fiducial
& 0.10
& 1.3
& \makecell{0.11--5.5\\(0.76)}
& \makecell{0.021--0.27\\(0.051)}
& \makecell{0.027--5.6\\(0.48)}
& \makecell{0.014--0.34\\(0.071)}
& 12
& 104
& 2--5 \\
\addlinespace[10pt]

Mdot3.8
& 0.10
& 3.8
& \makecell{0.089--11\\(1.6)}
& \makecell{0.037--0.57\\(0.11)}
& \makecell{0.16--10\\(2.0)}
& \makecell{0.045--0.41\\(0.20)}
& 12
& 57, 83
& 2--4 \\
\addlinespace[10pt]

Mdot0.56
& 0.10
& 0.56
& \makecell{0.14--5.0\\(0.51)}
& \makecell{0.017--0.21\\(0.035)}
& \makecell{0.011--2.1\\(0.10)}
& \makecell{0.010--0.26\\(0.041)}
& 8
& 141
& 2--3 \\
\addlinespace[10pt]

Mdot0.32
& 0.10
& 0.32
& 0.32
& 0.021
& 0.0076
& 0.0080
& \multicolumn{3}{c}{---} \\
\addlinespace[10pt]

Alpha0.06
& 0.06
& 1.3
& \makecell{0.14--5.2\\(0.75)}
& \makecell{0.023--0.25\\(0.050)}
& \makecell{0.043--4.6\\(0.46)}
& \makecell{0.022--0.33\\(0.084)}
& 18
& 130
& 5--7\\

\addlinespace[3pt]
\bottomrule
\end{tabular}%
}
\caption{Table~\ref{tablefull} summarizes the physical parameters for each model, including the mass supply rate at $100r_{\mathrm{s}}$ ($\dot{M}_{\mathrm{sup}}$), the mass accretion rate onto the black hole ($\dot{M}_{\mathrm{BH}}$), the disk luminosity ($L_{\mathrm{bol}}$), the mass outflow rate of the disk wind ($\dot{M}_{\mathrm{out}}$), and 
the fraction of solid angle over which the gas satisfied UFO criterion (covering factor). 
For the time-variable quantities ($\dot{M}_{\mathrm{BH}}$, $L_{\mathrm{bol}}$, $\dot{M}_{\mathrm{out}}$, and covering factor), the table lists the minimum and maximum values along with their time-averaged values in the format “minimum--maximum (average)”. Here, $t_{\mathrm{high}}$ is the duration of the high-luminosity phase,
$P_{\mathrm{cyc}}$ is the period of the luminosity cycle, and
$t_{\mathrm{delay}}$ is the time delay between the peaks of the total disk luminosity and the mass outflow rate.
Alt text: 
Table summarizing input parameters and simulation results for each model. Columns include the mass supply rate, black-hole accretion rate, disk luminosity, wind mass outflow rate, 
the 
solid angle of gas satisfying the UFO criterion,
and the duration, period of the luminosity cycle and time delay.}

\label{tablefull}
\end{table*}

\begin{figure}[htbp]
    \begin{center}
    \includegraphics[width=0.38\textwidth]{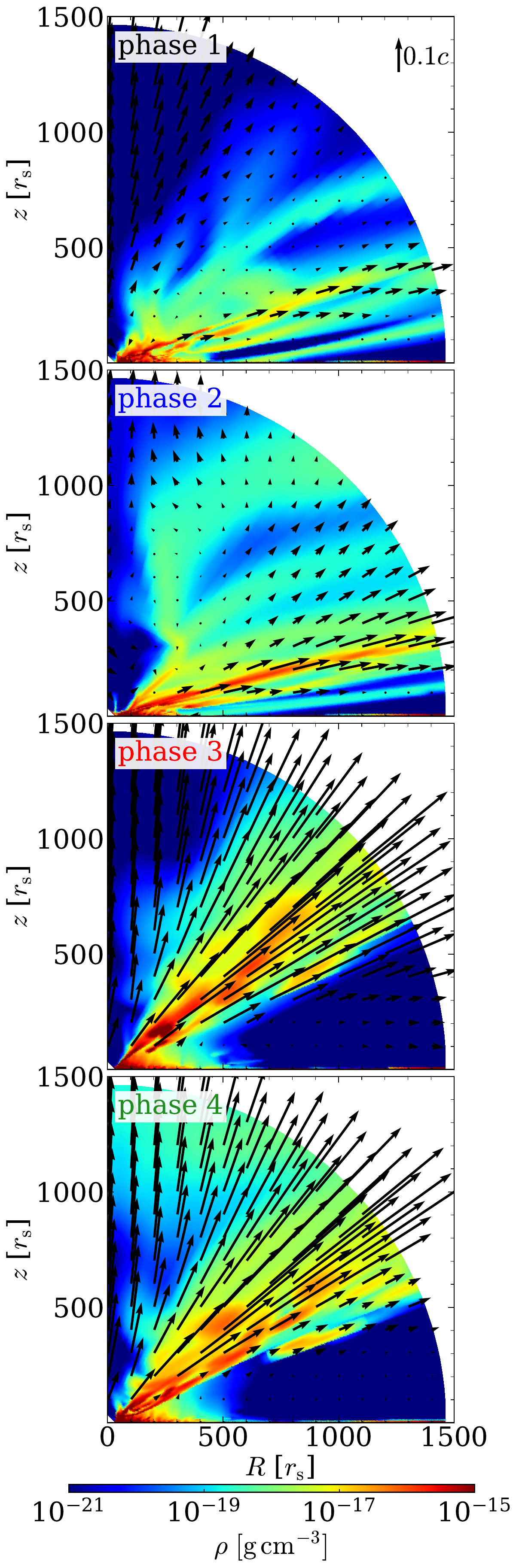} 
    \end{center}
    \caption{
    The density distribution and velocity vectors of the line-driven disk wind are shown for each phase in the cycle. From top to bottom, the panels correspond to phase 1, phase 2, phase 3, and phase 4. These results are obtained for the fiducial model. 
    Alt text:
    Four panels showing two-dimensional maps of disk-wind density over radial distance R and height z, with velocity vectors indicating the local wind direction and relative magnitude at four phases of the cycle.
    }
    \label{fig:wind4} 
\end{figure}

Figure \ref{fig:wind4} shows the structure of the disk wind at four phases.
The color contours show the density distribution, while the arrows indicate the velocity field of the wind.
In phases 1 and 2, the gas outflows are directed toward angles of approximately $75^\circ$ from the rotation axis. In contrast, during phases 3 and 4, the outflows shift to around $50^\circ$ and become both denser and faster. Specifically, in phase 1, the accretion rate, mass outflow rate, and covering factor are $0.12L_{\rm Edd}/c^2$, $0.053L_{\rm Edd}/c^2$, and 0, respectively. In phase 2, they are $4.7L_{\rm Edd}/c^2$, $0.65L_{\rm Edd}/c^2$, and 0.08. In phase 3, the values rise to $5.2L_{\rm Edd}/c^2$, $4.7L_{\rm Edd}/c^2$, and 0.32. In phase 4, the values are $0.98L_{\rm Edd}/c^2$, $3.5L_{\rm Edd}/c^2$, and 0.25.
Despite the luminosities 
in phases 2 and 4 being comparable (Fig.~\ref{fig:time_fidu}), the structure of the disk wind differs markedly between them. Both the mass outflow rate and the covering factor are higher in phase 4 than in phase 2. The cause of this difference will be discussed in a later section.

\begin{figure}[htbp]
    \begin{center}
    \includegraphics[width=0.5\textwidth]{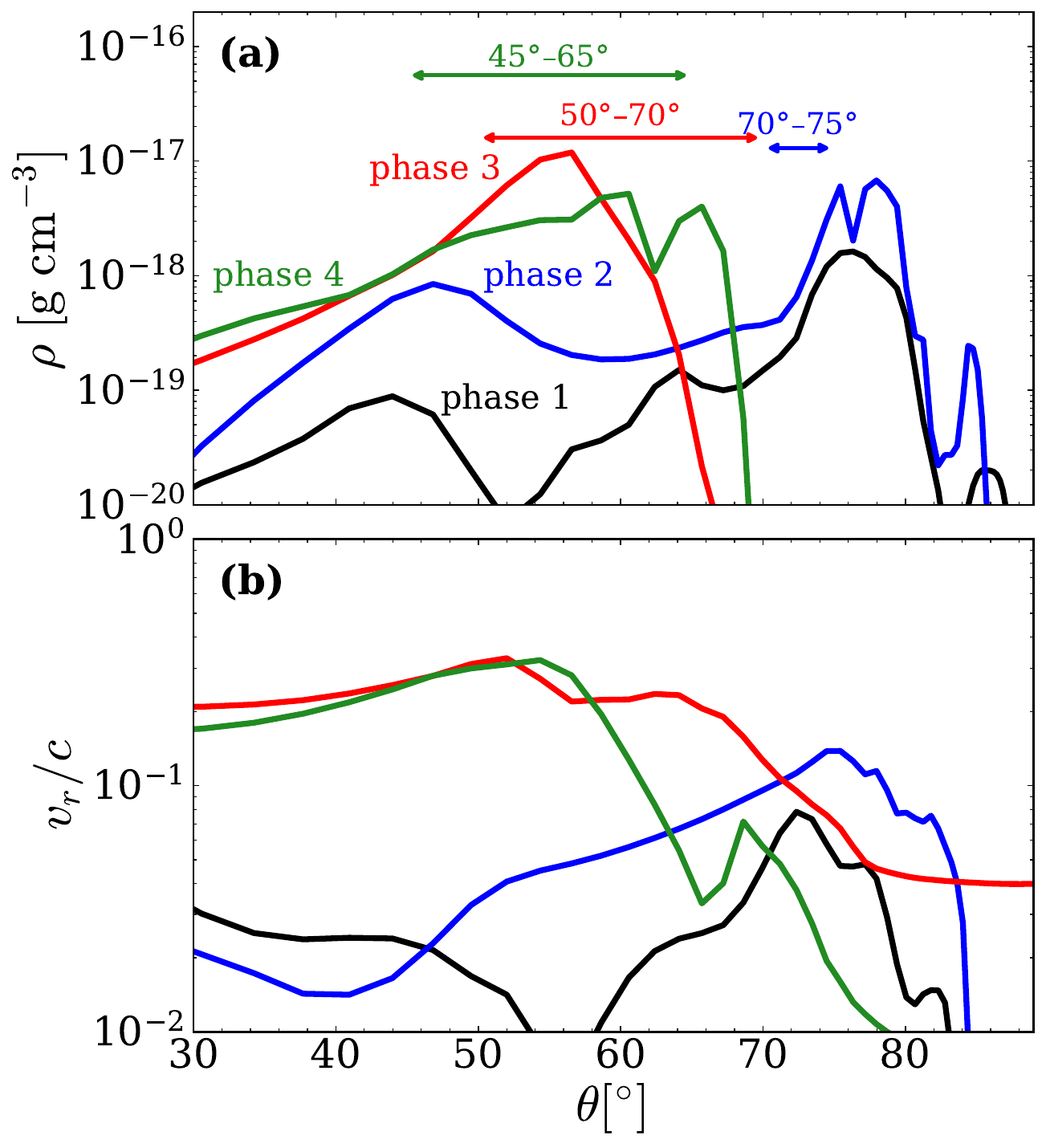} 
    \end{center}
    \caption{Polar angle distributions of the density and the regions satisfying the UFO criterion (panel a), and the radial velocity (panel b) for the fiducial model. 
    The black, blue, red, and green lines represent the results for phases 1, 2, 3, and 4, respectively. 
    Alt text: 
    Two line graphs labeled a and b showing angular profiles of density and radial velocity at four phases of the cycle. Each graph plots the quantities as functions of the polar angle, with four lines corresponding to the different phases.
    }
    \label{fig:wind_theta_4} 
\end{figure}

Figure \ref{fig:wind_theta_4} presents the angular distributions of the density and the radial component of the velocity at the outer boundary ($r=1500r_{\mathrm{s}}$) for the four phases. The peak density and velocity in each phase are approximately $2 \times 10^{-18}\mathrm{g~cm^{-3}}$ and $0.07c$ for phase 1, $8 \times 10^{-18}\mathrm{g~cm^{-3}}$ and $0.1c$ for phase 2, $10^{-17}\mathrm{g~cm^{-3}}$ and $0.3c$ for phase 3, and $8 \times 10^{-18}\mathrm{g~cm^{-3}}$ and $0.3c$ for phase 4.
In all phases, the angle at which the density reaches its maximum also corresponds to that 
of the maximum velocity. These directions coincide with the main outflow direction shown in Figure \ref{fig:wind4}. The angular ranges satisfying the UFO criterion are $70^\circ$ – $75^\circ$ for phase 2, $50^\circ$ – $70^\circ$ for phase 3, and $45^\circ$ – $65^\circ$ for phase 4 (see top panel and table \ref{tab:phases}), indicating that 
the gas satisfied by UFO criterion appear 
near the angles where both the density and velocity reach their maxima. This criterion is defined by the velocity, ionization parameter, and column density, as described in the text accompanying panel d of Figure 1. 
The resulting 
covering factors are 0.08, 0.32, and 0.25 for phases 2, 3, and 4, respectively. In phase 1, no angular region meets 
the UFO criterion, 
mainly due to insufficient velocity.
Our results imply that, depending on the viewing angle, UFOs may not appear even at high luminosities, whereas they may appear even at
low luminosities.

\begin{table}[htbp]
\centering
\begin{tabular}{ccccc}
\toprule
phase & $45^\circ$ – $50^\circ$ & $50^\circ$ – $65^\circ$& $65^\circ$ – $70^\circ$& $70^\circ$ – $75^\circ$ \\
\midrule
phase 1 & No & No & No  & No\\
phase 2 & No & No & No  & Yes\\
phase 3 & No & Yes & Yes  & No\\
phase 4 & Yes & Yes & No  & No\\
\bottomrule
\end{tabular}
\caption{
Viewing-angle ranges satisfying the UFO criterion for each phase of the fiducial model. Alt text: Table summarizing the viewing-angle ranges that satisfy the UFO criterion for each phase of the fiducial model. }
\label{tab:phases}
\end{table}

\begin{figure}[htbp]
    \begin{center}
    \includegraphics[width=0.38\textwidth]{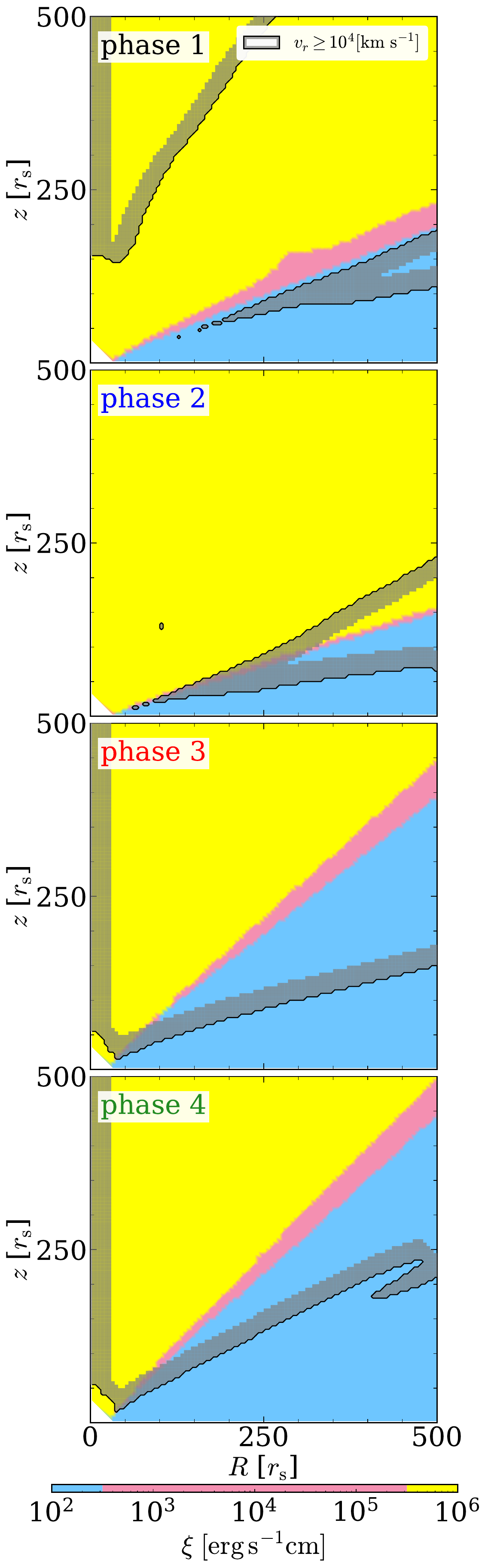} 
    \end{center}
    \caption{
Two-dimensional distributions of the ionization parameter in the fiducial model. Yellow, pink, and cyan indicate regions with $\log\xi > 5.5$, $\log\xi = 2.5$--$5.5$, and $\log\xi < 2.5$, respectively. The regions where the gas radial velocity satisfies $v_r > 10^4\ {\rm km\ s^{-1}}$ are enclosed by shaded black contours. Alt text: Two-dimensional ionization-parameter distributions in the fiducial model, with colors indicating three $\log \xi$ ranges and shaded black contours marking gas with $v_r > 10^4\ {\rm km\ s^{-1}}$.
    }
    \label{fig:xi} 
\end{figure}

Figure~\ref{fig:xi} shows the two-dimensional distributions of the ionization parameter and radial velocity on the $R-z$ plane in each phase. 
The range of ionization
parameters required to satisfy the UFO criterion,
$2.5 \le \log \xi \le 5.5$, is shown in pink, while regions with radial
velocities in the range required to satisfy the UFO criterion,
$v_r > 10^4~{\rm km~s^{-1}}$, are enclosed by black shaded solid contours.
In phase 2, a region satisfying both the ionization-parameter range
($2.5 \le \log \xi \le 5.5$) and the radial-velocity range
($v_r > 10^4~{\rm km~s^{-1}}$) appears at
$r \gtrsim 100r_{\rm S}$ along viewing angles of $\sim 70^\circ$.
In contrast, in phases 3 and 4, regions satisfying both ranges extend
along viewing angles of $\sim 50^\circ$--$60^\circ$ at
$r \gtrsim 100r_{\rm S}$.
In phase 1, no region satisfies both ranges. Although regions with
$\log \xi < 2.5$ are present, their velocities do not fall within the
required radial-velocity range. Conversely, the ionization parameters in
regions with $v_r > 10^4~{\rm km~s^{-1}}$ do not fall within the required
ionization-parameter range,
$2.5 \le \log \xi \le 5.5$.

\begin{figure}[htbp]
    \begin{center}
    \includegraphics[width=0.5\textwidth]{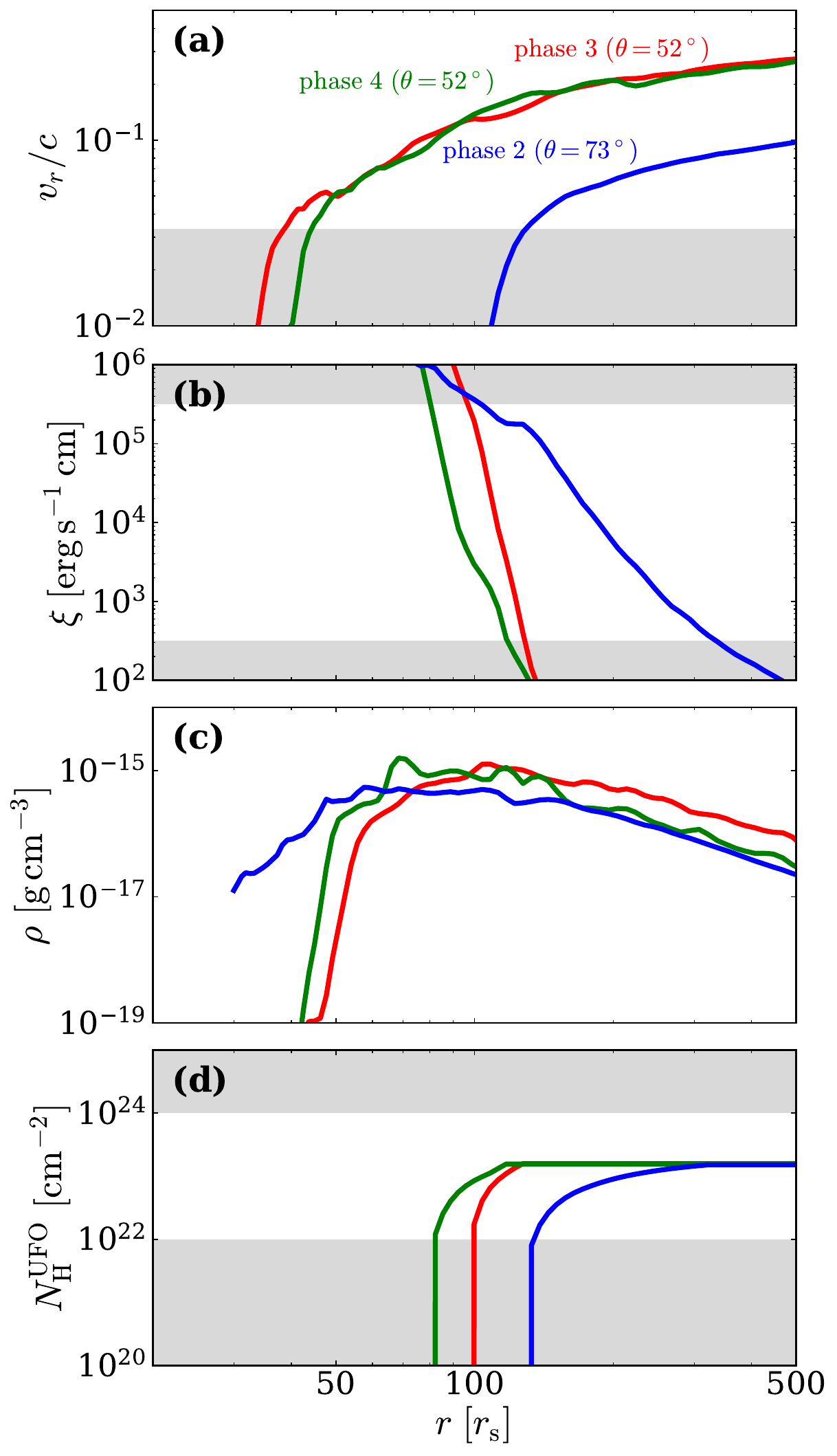} 
    \end{center}
    \caption{
    Radial profiles of the radial velocity (panel a), ionization parameter (panel b), and density (panel c) along representative viewing angles at which gas satisfying the UFO criterion is detected in the fiducial model. Panel d shows the column density of gas with velocities higher than $10^4\ {\rm km\ s^{-1}}$ and ionization parameters in the range $\log\xi = 2.5$--$5.5$. The red, blue, and green curves correspond to phase 3 at a viewing angle of $52^\circ$, phase 2 at a viewing angle of $73^\circ$, and phase 4 at a viewing angle of $52^\circ$, respectively. The gray regions in panels a, b, and d indicate where the corresponding physical quantities fall outside the ranges required to satisfy the UFO criterion. Alt text: Radial profiles of radial velocity, ionization parameter, density, and column density of gas with velocities higher than $10^4\ {\rm km\ s^{-1}}$ and ionization parameters in the range $\log\xi = 2.5$--$5.5$ at representative viewing angles in the fiducial model. The colored curves represent different phases and viewing angles. Gray regions indicate values outside the UFO criterion.
    }
    \label{fig:r_profile} 
\end{figure}

A more precise location of the regions satisfying the UFO criterion can be inferred from Figure~\ref{fig:r_profile}. This figure shows the radial velocity (panel a), ionization parameter (panel b), and density (panel c) along representative viewing angles where the UFO criterion is satisfied. Panel d shows the column density of gas with velocities $>10^4\ {\rm km\ s^{-1}}$ and ionization parameters in the range $2.5 \leq \log\xi \leq 5.5$.
The gray regions in panels a, b, and d indicate where the corresponding physical quantities fall outside the values required to satisfy the UFO criterion. 
As seen in Figure~\ref{fig:r_profile}, the radial velocity falls within
the range required to satisfy the UFO criterion at radii larger than
$\sim 100r_{\rm s}$ in phase 2, and at $r\gtrsim 40r_{\rm s}$ in
phases 3 and 4. However, the ionization parameter falls within the
required range only within $r \sim 100-400r_{\rm s}$ in phase 2 and
$r \sim 80-150r_{\rm s}$ in phases 3 and 4.
Therefore, the gas responsible for the blueshifted iron absorption lines
characteristic of UFOs is located at around $100r_{\rm s}$.

\subsection{Influence of effective temperature}

Although the disk luminosities are similar, the disk-wind structure differs markedly between phases 2 and 4, 
as shown 
in Figs. \ref{fig:wind4} and \ref{fig:wind_theta_4}. In phase 4, the wind is launched at higher velocities and directed toward higher latitudes, farther from the disk plane. In contrast, the wind in phase 2 is slower and launched closer to the disk plane. Both the mass outflow rate and the covering factor are larger  
in phase 4 than in phase 2. These differences originate from variations in the effective temperature of the disk.

As shown in Figure 2, although the effective temperature in the innermost region is higher in phase 2, the temperature in the region close to the wind base
is higher in phase 4. The higher $T_{\rm eff}$ near the wind base enhances the vertical component of the radiation force from the disk, thereby launching stronger disk winds away from the disk plane. Consequently, the mass outflow rate in phase 4 exceeds that in phase 2, and the polar angles at which UFOs are detected become smaller in phase 4. The angular range satisfying UFO criterion also broadens, and thus the covering factor becomes large  
in phase 4.
Here, the wind base refers to the region extending approximately $10r_{\mathrm{s}}$ around $R_{\mathrm{launch}}$, as defined in Equation \eqref{eq:16}, from which the line-driven disk wind is mainly launched. In the present model, $R_{\mathrm{launch}}$ is approximately 
$77r_{\mathrm{s}}$, $64r_{\mathrm{s}}$, $40r_{\mathrm{s}}$, and $41r_{\mathrm{s}}$
for phases 1--4, respectively.

The outflow direction of the disk wind in phase 4 is similar to that in phase 3 since 
the effective temperatures near the 
wind base 
are comparable. However, in phase 3 the innermost disk temperature is higher, which strengthens 
the radial component of the radiation force and tends to produce the high velocity winds 
(see $\theta \gtrsim 65^\circ$ in Fig. \ref{fig:wind_theta_4}). Comparing phases 2 and 1, the winds have similar morphology, as 
the effective temperature near the wind base is similar. The stronger wind in phase 2 arises from the higher temperature in the region closer to the black hole.

Accordingly, the strength and structure of the disk wind vary in
response to changes in the effective-temperature profile, causing
the variation in the mass outflow rate to lag 
slightly
behind that
of the disk luminosity (see also Fig.~1). 
We suggest that this
time delay is caused by the viscous evolution around the wind base. Indeed, the
viscous timescale is estimated as
$t_{\rm vis}\sim 2.8
(M_{\rm BH}/10^{7.4}M_\odot)
(R/40r_{\rm s})^{3/2}
(\alpha/0.1)^{-1}
(H/R/0.1)^{-2}\ {\rm yr}$
(Pringle 1981). Here, $R\sim40r_{\rm s}$ corresponds to the characteristic
wind-base radius during phases 3 and 4, when the mass outflow rate is relatively
high. For $\alpha=0.1$ and $H/R=0.1$, the viscous timescale is approximately
3 yr, which is comparable to the 2--5 yr time delays listed in Table~\ref{tablefull}.
For the model with $\alpha=0.06$, the calculated time delay is also comparable
to the corresponding viscous timescale.
In contrast, the instability-front propagation timescale (Hameury et al. 2009), $(H/R)t_{\rm vis}$, 
is about $0.3\,{\rm yr}$ for $H/R \sim 0.1$
and the outflow-propagation timescale, $100r_{\rm s}/0.1c$, is estimated as $\sim 8\times10^{-3}\,{\rm yr}$.
Both timescales are much shorter than the time delays of a few years obtained
in our simulations. Therefore, these propagation timescales cannot account for
the time delays, suggesting that the viscous evolution around the wind base
determines the time delay.

It should be noted that while the limit-cycle oscillation also induces 
density variations in the disk, the structure of the disk wind is predominantly 
determined by the effective temperature. Indeed, even if the disk density (more precisely, the density boundary condition near the disk surface in the wind simulation) is fixed, variations 
in the effective temperature still alter the wind structure. On the other hand, if $T_{\rm eff}$ is fixed, varying the disk density has only a minor 
effect on the wind structure.

\begin{figure}[htbp]
  \begin{center}
    \includegraphics[width=0.5\textwidth]{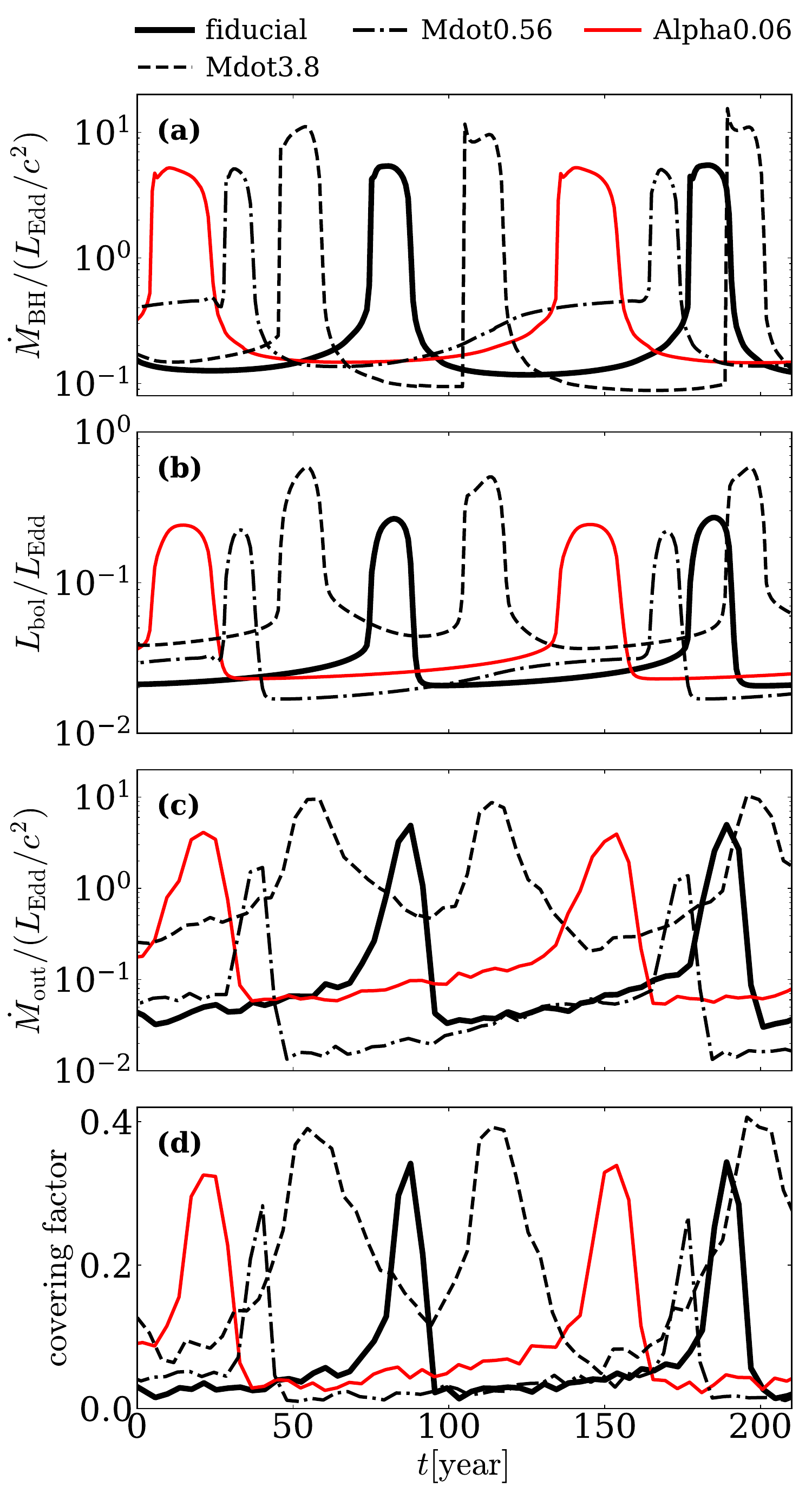}

  \end{center}
    \caption{Time evolution of the mass accretion rate onto the black hole (panel a), the disk luminosity (panel b), the mass outflow rate of the disk wind at $1500r_{\mathrm{s}}$ (panel c), and the covering factor of gas satisfying the UFO criterion (panel d). The black solid line represents the fiducial model, the black dashed line the Mdot3.8 model, the black dash-dotted line the Mdot0.56 model, 
    and the red solid line the Alpha0.06 model. Alt text:
    Multi-panel figure with four line graphs showing, from panels a to d, the time evolution of the accretion rate, disk luminosity, mass outflow rate, and covering factor. Several models are distinguished by different line styles and colors. 
    }
    \label{fig:time} 
\end{figure}

\subsection{Parameter dependences}
In addition to the fiducial model, we adopt models with different values of $\alpha$ and $\dot{M}_{\rm sup}$. 
A summary of the results is provided in Table \ref{tablefull}.

In addition to the fiducial model (black solid line, $\dot{M}_{\mathrm{sup}}=1.3L_{\mathrm{Edd}}/c^2$), 
Figure \ref{fig:time} shows the Mdot3.8 model (dashed line) with a larger mass supply rate, $\dot{M}_{\mathrm{sup}}=3.8L_{\mathrm{Edd}}/c^2$, and the Mdot0.56 model (dash–dotted line) with a smaller supply rate, $\dot{M}_{\mathrm{sup}}=0.56L_{\mathrm{Edd}}/c^2$. In both models, as in the fiducial model, the luminosity rises and falls in step with changes in the mass accretion rate. The oscillation period is nearly the same for the fiducial model and the Mdot3.8 model, but 
both the accretion rate and the luminosity are systematically higher in the Mdot3.8 model. In contrast, for the Mdot0.56 model, the maximum and minimum luminosities are comparable to those in the fiducial model, and 
the period is longer. Consequently, the time-averaged accretion rate and luminosity are smaller in this model than in the fiducial model. The mass outflow rate 
increases with the mass supply rate, and the 
covering factor follows the same trend. 
The maximum and time-averaged 
covering factors are 0.41 and 0.20 for the Mdot3.8 model, 0.34 and 0.071 for the fiducial model, and 0.26 and 0.041 for the Mdot0.56 model, respectively. 
More detailed values are summarized in Table \ref{tablefull}. For an even lower supply rate (Mdot0.32 model), the limit-cycle oscillation does not occur.

When a smaller viscosity parameter of $\alpha = 0.06$ is adopted (compared to $\alpha = 0.1$ in the fiducial model), the period increases to approximately 130 yr, which is about 1.6 times longer than that of the fiducial model. The duration of the high-luminosity state is also extended to about 18 yr, approximately 1.5 times longer than that in the fiducial case. These extensions in both period and duration are due to the increase in the viscous timescale. In fact, the viscous timescale for $\alpha = 0.06$ is approximately 1.6 times longer than that in the fiducial model. The covering factor, however, is nearly the same as that in the fiducial model.

\section{Summary and Discussions}
In this study, we simultaneously perform one-dimensional hydrodynamic simulations of an accretion disk and two-dimensional radiation hydrodynamic simulations of a line-driven disk wind, in order to investigate the self-consistent structure and evolution of both components. Our results demonstrate that disk luminosity variations caused by the 
radiation pressure instability in the disk drive time variability in line-driven disk winds, which lead to the episodic appearance of 
the gas satisfied by UFO criterion. The disk 
instability arises when the mass supply rate (the mass accretion rate at $r \sim 100r_{\rm s}$) is around $L_{\mathrm{Edd}}/c^2$, 
while it does not occur when the rate is approximately less than $0.3L_{\mathrm{Edd}}/c^2$. 
Therefore, luminosity variations and episodic appearances of 
the gas satisfied by UFO criterion observed in AGNs with luminosities around 10\% of the Eddington luminosity can be explained by this mechanism.

When the disk luminosity increases (or decreases) due to 
radiation pressure instability, the mass outflow rate driven by the line force also increases (or decreases). However, the variation in mass outflow rate lags
slightly 
behind the luminosity variation by a timescale comparable to 
the viscous timescale around the wind base. 
This delay can be understood as arising from the fact that
the mass outflow rate is primarily
determined by the effective-temperature profile near the wind base
rather than by the total disk luminosity. 
The effective-temperature profile near the wind base evolves differently from the total disk luminosity, causing the mass outflow rate to respond with a slight delay to the luminosity variation.
This is 
because during the luminosity rising phase, the effective temperature near the inner edge of the disk increases, whereas during the declining phase, the temperature becomes higher around the wind base region located several tens of Schwarzschild radii from the black hole. 
A higher effective temperature (stronger radiation flux) around the wind base promotes efficient disk wind launching. As a result, the mass outflow rate becomes largest during the transition from the high luminosity phase to the declining phase. 
The covering factor of viewing angles satisfying the UFO criterion 
(the column density of gas with $2.5 \le \log \xi \le 5.5$ and $v_r > 10^4~{\rm km~s^{-1}}$ is $N_{\rm H}=10^{22}$--$10^{24}~{\rm cm^{-2}}$) 
 follows the variation in mass outflow rate and therefore also lags slightly behind the  
luminosity variations. In the case where the mass supply rate is $1.3L_{\mathrm{Edd}}/c^2$, the viscosity parameter is $\alpha = 0.1$, and the viscosity prescription parameter is $\mu = 0.45$ (the fiducial model), the covering factor reaches approximately 30\% during the high luminosity and declining phases, but remains only a few percent during other phases.

The direction in which the disk wind is launched and the viewing angle at which 
UFO criterion is satisfied 
both vary with time. During the rising phase of luminosity, the wind is mainly launched toward relatively edge-on directions, and 
UFO criterion also tends to be satisfied 
along such lines of sight. During the high-luminosity phase, 
the opening angle of the disk wind becomes narrower, making the viewing angles at which UFO 
criterion can be satisfied smaller (i.e., more face-on). 
The smallest 
angle satisfied by UFO criterion (the most face-on) occurs during the declining phase. For example, in the fiducial model, the 
angles satisfied by UFO criterion
are $70^{\circ}-75^{\circ}$ during the rising phase, $50^{\circ}-70^{\circ}$ during the high-luminosity phase, and $45^{\circ}-65^{\circ}$ during the declining phase. 
In other words, this means that the presence or absence of UFOs varies with time and that, depending on the viewing angle, UFOs may not appear even at high luminosities, whereas they may appear even at
low luminosities.

Our results can explain the observational fact that the presence or absence of UFOs is not necessarily correlated with luminosity. 
For example, in 2MASS~0918+2117, no UFO was detected during an epoch with $L_{\rm bol}/L_{\rm Edd} \sim 0.44$, whereas a UFO was detected during an epoch with $L_{\rm bol}/L_{\rm Edd} \sim 0.11$. In addition, there was also an epoch with a comparable luminosity ($L_{\rm bol}/L_{\rm Edd} \sim 0.13$) in which no UFO was detected \citep{Baldini2024}.
In PG1448+273, UFOs were not detected during epochs with $2$--$10$ keV luminosities of $\sim 2.2\times10^{43} \mathrm{erg\ s^{-1}}$ and $\sim 4.8\times10^{43} \mathrm{erg\ s^{-1}}$, whereas UFOs were detected during epochs with luminosities of $\sim 1.8\times10^{43} \mathrm{erg\ s^{-1}}$ and $\sim 2.6\times10^{43} \mathrm{erg\ s^{-1}}$. Thus, there is no clear correlation between UFO detection and luminosity \citep{Kosec2020,Reeves2023,Reeves2024}. Furthermore, a similar trend, in which UFOs are detected in some epochs but not in others even at comparable luminosities, has also been found in PG1115+080 and HS~0810+2554 \citep{Chartas2021}.
In addition, for the Seyfert galaxies PG1211+143, Mrk509, Mrk766, Mrk841, Mrk79, and NGC4151, it has been reported that UFO detections and non-detections cannot be distinguished solely by the X-ray flux \citep{Tombesi2010}.
In this study, we investigated the viewing angles at which UFOs can be detected based on the velocity, ionization parameter, and column density. However, more direct comparisons with observations require radiative transfer calculations and synthetic X-ray spectra \citep{Schurch2009,Sim2010,Mizumoto2021}. Mizumoto et al. (2020) reproduced absorption spectra using simulation results of line-driven disk winds launched from a steady disk. Reproducing the absorption-line structures produced by disk winds that vary with time because of disk 
radiation pressure instability, and comparing them with observational data, is important for future work.

The disk winds obtained in our simulations are likely capable of explaining UFOs, whereas reproducing Broad Absorption Lines (BALs), Narrow Absorption Lines (NALs), and Warm Absorbers (WAs) would be difficult. This is because their velocities are too high compared with those of NALs and WAs, while their velocity dispersion is not large enough to reproduce BALs.
However, this discussion is limited to the computational domain considered here. If the disk winds interact with the surrounding interstellar medium at larger radii and density and velocity structures of the wind change substantially, they may still form diverse absorption structures. Understanding the evolution of disk winds at larger radii 
is also important for future work.

In this study, we solve the accretion disk and the line-driven disk wind simultaneously while accounting for their mutual interaction. However, there are several important physical processes that are not fully treated 
in this approach. For instance, some of the wind material may fail to escape and eventually fall back onto the disk, affecting the disk structure. This effect is 
not included in the present  
disk simulations. Additionally, emission  
and scattering from the disk wind material could affect the temperature distribution on the disk surface.
But this feedback is also neglected. 
To resolve these issues, it is necessary to conduct large-scale radiation hydrodynamic simulations that resolve both the accretion disk and the disk wind within a single computational domain, 
rather than treating them separately as in this study. Such simulations are left for future work.
Moreover, the present study does not account for magnetic fields. Previous studies have shown that magnetic fields can drive disk winds, and they can also 
significantly alter the structure of line-driven winds \citep{Proga2003, Yang2021}. Simulations including magnetic effects also remain for future investigation.
While we explored several models by varying the mass accretion rate, viscosity parameter $\alpha$, and viscosity prescription parameter $\mu$ 
around the fiducial model, a wider 
parameter survey is necessary to constrain the allowable range of parameters. 
Because of computational limitations, it is currently difficult to perform the long-term simulations required in this study, namely simulations extending over $\gtrsim 10^7 (GM_{\rm BH}/c^3)$. Nevertheless, future studies should relax the assumptions of the standard accretion-disk model and treat the disk–wind system more self-consistently using radiation magnetohydrodynamic simulations \citep{Blaes2025}. Achieving such long-term radiation magnetohydrodynamic simulations remains an important challenge for future work.

This study still requires several improvements for a more realistic treatment. One of which is a more accurate modeling of the line-driving force.
In this study, we adopted a conventional approximation used in earlier works \citep{Stevens1990}. However, previous studies have pointed out that reprocessed photons can extend the ionized regions, weaken the line force, and make the disk winds less powerful
\citep{Higginbottom2014, Higginbottom2024, Mosallanezhad2025}. It has also been reported that 
X-ray scattering and re-emission induce time variability in line-driven disk winds \citep{Dyda2024}.
On the other hands, \citet{Dannen2019} and \citet{Dyda2025} found that the line force, when spectral lines in the X-ray band previously neglected are taken into account, is enhanced and the mass outflow rate of the line-driven wind increases significantly. Furthermore, radiation drag, which is not considered in the present study, may reduce the mass outflow rate by several tens of percent according to \citet{Wang2022} and \citet{Tang2025}.
Because recent studies suggest that UFOs may have clumpy structures \citep{Audard2025,Xu2025,Xiang2025,Mehdipour2025}, simulations without the assumption of axisymmetry are also required. Although 
in the case of disk winds around white dwarfs, three-dimensional radiation hydrodynamic simulations have successfully reproduced the formation of numerous gas clumps in line-driven winds \citep{Dyda2018}.

\section*{Acknowledgments}
We would like to thank Takumi Ogawa and Yuta Asahina for useful discussions. The numerical simulations were performed on HPE Cray XD2000 at the Center for Computational Astrophysics (CfCA), National Astronomical Observatory of Japan. This work was also supported (in part) by the Multidisciplinary Cooperative Research Program in CCS, University of Tsukuba.

\section*{Funding}
Y.K. was supported by Japan Science and Technology Agency (JST) Support for Pioneering Research Initiated by the Next Generation (SPRING), Japan Grant Number JPMJSP2124. This research was supported by the Japan Society for the Promotion of Science (JSPS) through KAKENHI Grant Numbers JP21H04488 (KO), 24K00678 (KO), 25K01045 (KO), and JP26K07146 (MN).
This work was also supported by MEXT as “Program for Promoting Researches on the Supercomputer Fugaku” (Structure and Evolution of the Universe Unraveled by Fusion of Simulation and AI; Grant Number JPMXP1020240219; KO), by Joint Institute for Computational Fundamental Science (JICFuS, KO), and by the Exploratory Research Grant for Young Scientists, Hirosaki University (MN).

\end{document}